\documentclass[twocolumn,showpacs,preprintnumbers,amsmath,amssymb,superscriptaddress]{revtex4}
\usepackage{graphicx}
\usepackage{dcolumn}
\usepackage{bm}

\begin{document}


\title{Degree distributions in mesoscopic and macroscopic functional brain networks}

\author{Satoru Hayasaka}
\affiliation{Department of Biostatistical Sciences,
Wake Forest University Health Sciences,
Winston--Salem, North Carolina, 27157, USA
}
\affiliation{Department of Radiology,
Wake Forest University Health Sciences,
Winston--Salem, North Carolina, 27157, USA
}
\author{Paul J. Laurienti}
\affiliation{Department of Radiology,
Wake Forest University Health Sciences,
Winston--Salem, North Carolina, 27157, USA
}

\date{\today}

\begin{abstract}
We investigated the degree distribution of brain networks extracted from functional magnetic
resonance imaging of the human brain. 
In particular, the distributions are compared between macroscopic
brain networks using region-based nodes and mesoscopic brain networks using voxel-based
nodes. We found that the distribution from these networks follow the same family of distributions
and represent a continuum of 
exponentially truncated power law distributions. 
\end{abstract}

\maketitle

%
%
Small-world networks are a class of networks characterized by highly interconnected 
neighborhoods and efficient long-distance connections, connecting any two nodes in a
network with just a few intermediary connections  \cite{watts:strogatz}. 
Since the introduction of small-world networks, these small-world
properties have been observed in many social, technological, and biological networks
\cite{strogatz:review}. The network organization of the human brain has also been 
demonstrated as a small-world network \cite{stam:review,bullmore:review};
small-world properties have been
verified in both anatomical \cite{gong,he:cort,hagmann:core,iturria} and
functional \cite{achard:fmri,bassett:meg,cecchi,eguiluz,heuvel,stam:meg} brain networks.
A small-world structure is advantageous for brain networks since it can
support both localized processes specific to different brain regions as well as distributed
processes encompassing multiple brain regions at once. Some studies have also reported
that functional brain networks are scale-free networks \cite{cecchi,eguiluz,heuvel}, networks
characterized by mega-hubs with extremely large node degrees
and by the degree distribution following a power law distribution \cite{barabasi:scalefree}.

Interestingly, scale-free properties have been observed in brain networks using voxels 
(3D pixels in 3D brain images) as network nodes  \cite{cecchi,eguiluz,heuvel}, 
but not in brain networks modeled using anatomical 
regions as nodes \cite{achard:fmri,bassett:meg,gong,he:cort,iturria}. 
The distribution from a region-based network is not truly scale-free but
follows an exponentially truncated power law distribution \cite{achard:fmri,gong,he:cort}.
This discrepancy is rather surprising since both types of networks describe the same biological
system, the human brain. Although the reason for the difference is unclear, it may be due to
differences in data processing steps in constructing the brain network \cite{bullmore:review}. 
Another possible reason for the discrepancy is the difference in the scale of these networks;
while voxel-based networks represent a finer mesoscopic organization of the brain, 
region-based networks represent a coarser macroscopic organization of the brain.
Thus, to examine the effects of data processing and the network scale on the node degree
distribution, we constructed region-based and voxel-based networks based on the
same functional MRI (fMRI) data. We compared the degree distribution from both types of
networks describing the organization of the same set of brains.

%
%
\begin{figure}
\includegraphics{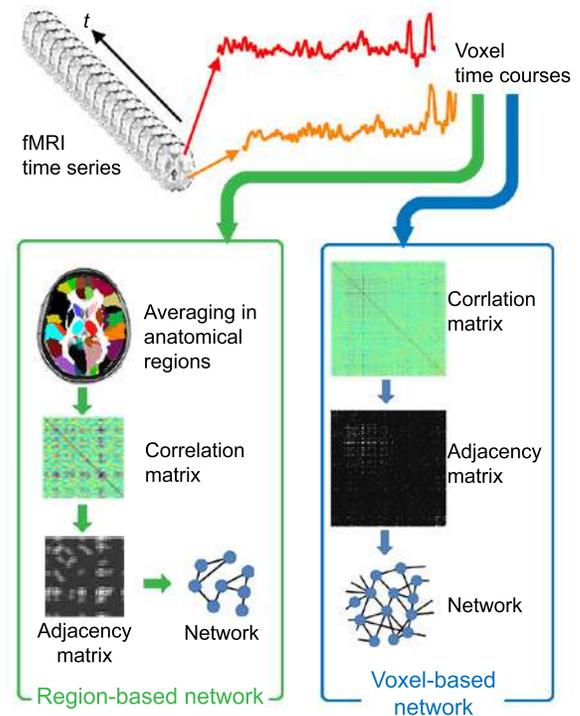}
\caption{A schematic of data processing steps to generate a region-based
network and a voxel-based network from an individual fMRI data set.
Voxel time courses are extracted from a series of 3D fMRI images.
In a region-based network, extracted time courses are averaged
in different anatomical regions to produce a node time course
in each region, whereas in a voxel-based network each voxel time course
is treated as a distinct node time course. 
A correlation matrix is generated between node
time courses, and then thresholded by a correlation threshold $R_c$
to produce a binary adjacency matrix representing a network. }
\label{fig:schematic}
\end{figure}

The data set for this study consisted of fMRI experiment data from 5 subjects. For each subject,
a series of MRI images measuring neurological activities were acquired every 2.5 sec. for 5 minutes,
resulting in 120 images, each with $46 \times 55 \times 37$ voxels of size $4\times 4\times 5$mm. 
FIG \ref{fig:schematic} displays the schematic of the data processing steps.
To construct a voxel-based network, each of $\sim$20,000 brain voxels was treated as a node.
For a region-based network, the voxel time courses were averaged for voxels within each anatomical
region, resulting in time courses corresponding to approximately 100 nodes of anatomical regions.
In both networks, the correlation coefficient was calculated between the time courses of two nodes, 
producing a correlation matrix. If the correlation coefficient exceeded a
correlation threshold $R_c$, these nodes were considered to be {\it functionally connected}.
By applying a threshold $R_c$ to the correlation matrix, a binary matrix was
formed, known as an adjacency matrix, with 1 indicating existence of an edge 
connecting two nodes and 0 otherwise.
To facilitate a comparison of degree distributions across subjects, 
$R_c$ was adjusted for each subject so that the average 
node degree for the network was similar across subjects.  The average degree was 
$30.3 \pm 1.3$(SD) for the voxel-based networks and $4.44 \pm 0.19$
for the region-based networks.

%
%
\begin{figure}
\includegraphics{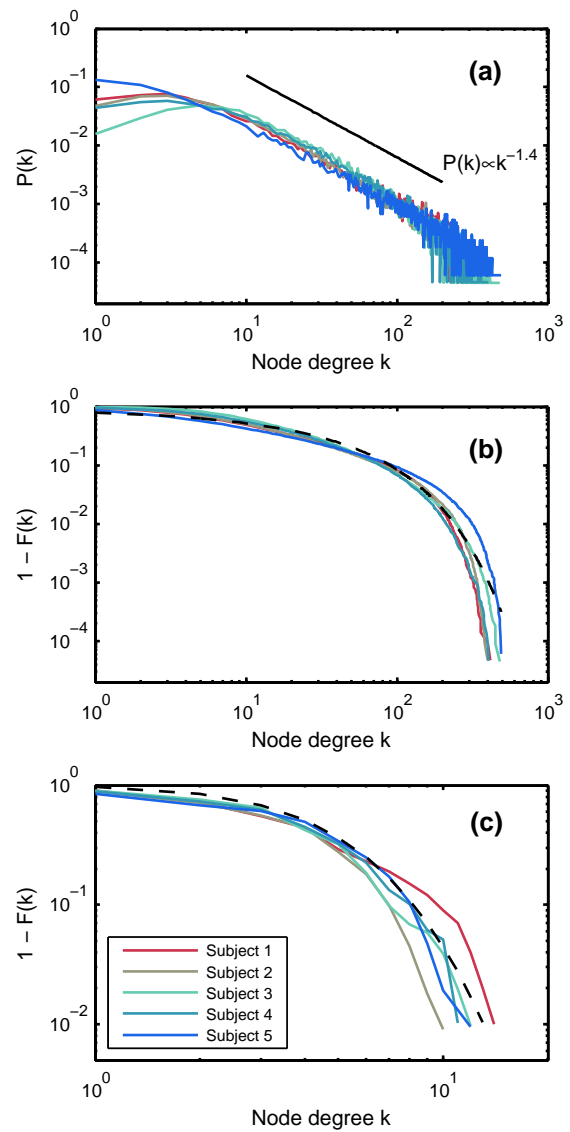}
\caption{Node degree distributions from the voxel-based and region-based networks:
the probability distributions $P(k)$ from the voxel-based networks (a), the complimentary 
cumulative distributions $1-F(k)$ from the voxel-based networks (b), and the complementary
cumulative distributions from the region-based networks (c). 
The best-fit curves of exponentially truncated power law 
distribution $P(k)\propto k^{\beta} \exp(-k/ k_c)$ (dashed curves) are also shown, 
with $\beta=-0.61$ and $k_c=81.3$ for the voxel-based networks (b) and
$\beta=1.87$ and $k_c=1.60$ for the region-based networks (c).}
\label{fig:distplot}
\end{figure}

FIG \ref{fig:distplot} show the degree distributions resulting from the voxel-based and region-based
networks. FIG \ref{fig:distplot}(a) shows the degree distributions of the voxel-based networks plotted on a
log--log scale. For all the subjects, the degree distribution $P(k)$ seems to follow a straight line 
as seen in other voxel-based networks \cite{cecchi,eguiluz,heuvel}, indicative
of a power law distribution $P(k)\propto k^{-\gamma}$ with $\gamma\simeq1.4$. 
However, the tail of the distributions 
exhibit increased uncertainty, giving an appearance of a {\it fuzzy} tail. 
To understand the distribution profile better \cite{keller},
we calculated the cumulative distribution $F(k) = \Sigma_{k' \leq k} P(k')$ and 
plotted the complementary cumulative distribution $1-F(k)$ on a log--log scale in FIG \ref{fig:distplot}(b). 
If the distribution were truly a power law distribution, then the plot of $1-F(k)$ would also follow 
a straight line \cite{keller}.  Instead, the distributions decay faster than a power law distribution,
and follow an exponentially truncated power law
distribution $P(k)\propto k^{\beta} \exp(-k/ k_c)$ with $\beta=-0.61$ and $k_c = 81.3$ 
(dashed curve, FIG \ref{fig:distplot}(b)). 

FIG \ref{fig:distplot}(c) shows the log-log plots of
$1-F(k)$ from the region-based networks. The distributions are curved, showing an accelerated
decay for higher $k$. These distributions follow a gamma distribution 
$P(k)\propto k^{\alpha-1} \exp(-k/ k_c)$, which is another parameterization of an
exponentially truncated power law distribution with the exponent $\alpha-1$ instead of 
$ \beta$. As a reference, FIG \ref{fig:distplot}(c) also shows the best-fit gamma distribution, 
with $\alpha=2.87$ and $k_c = 1.60$ (dashed curve). 
These parameters are within a similar range as other region-based functional and anatomical 
brain networks \cite{achard:fmri,bassett:meg,gong,he:cort,iturria}.

%
%
The results from both types of networks demonstrate that the degree distribution follows
an exponentially truncated power law distribution. This is not surprising since both networks
are modeling functional connectivities of the same brains, although scales are different.
As a comparison, we also fitted other distributions, namely a power law distribution 
$P(k)\propto k^{-\gamma}$ and an exponential distribution $P(k)\propto \exp(-k/k_c)$, 
but neither approximated the distribution better than an exponentially truncated power law distribution. 
It is interesting that the degree distributions $P(k)$ from voxel-based networks appear to follow a 
power law distribution as previously reported by other studies. However, the plots of the cumulative
distributions $1-F(k)$ indicate otherwise. Therefore the scale-free properties seen in other voxel-based
networks \cite{cecchi,eguiluz,heuvel} may simply be an artifact of data display. Since there are very few
nodes with high $k$, the empirical estimation of $P(k)$ based on a histogram is very unstable
as noted by Keller \cite{keller}, and a straight line can be erroneously fitted to other distributions
\cite{li:scalefree}.

Interestingly, the characteristics of the exponentially truncated power law distribution 
are different between the region-based and voxel-based networks. In particular,
in a voxel-based network, the degree distribution has a form $P(k)\propto k^{\beta} \exp(-k/ k_c)$
with $\beta<0$, whereas in a region-based network, $\beta>0$. In a voxel-based network, 
$k^{\beta}$ decreases as $k$ increases, and the decay is further accelerated by the exponential 
truncation $\exp(-k/k_c)$. On the other hand, in a region-based network, 
$k^{\beta}$ increases as $k$ increases, but this increase is attenuated and eventually 
overcome for $k\gg k_c$ by the exponential truncaton $\exp(-k/k_c)$.
If $\beta=0$, then the distribution is an exponential distribution
$P(k)\propto \exp(-k/k_c)$ characterized by a single cut-off parameter $k_c$. 

\begin{figure}
\includegraphics{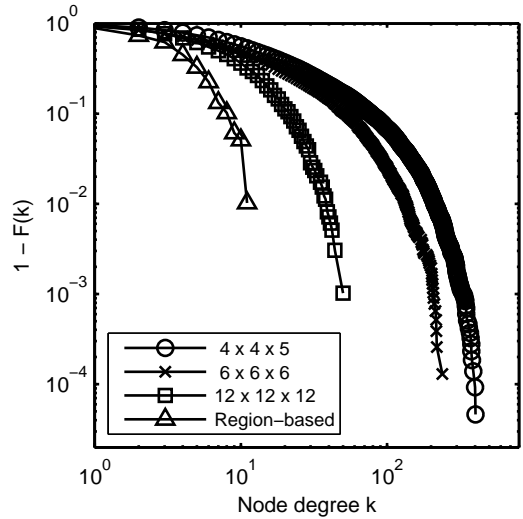}
\caption{Node degree distributions $1-F(k)$ of voxel-based networks of a single
subject generated with different voxel sizes. The degree distribution of the region-based
network for the same subject is also shown.}
\label{fig:scale}
\end{figure}

%
%
Since the sign of the exponent $\beta$ is the only distinction between the degree distributions 
of the two types of networks, it is plausible that the degree distribution of a brain network is
from a continuum of exponentially truncated power law distributions,
with $\beta$ dependent on the scale at which the network is formed.
To verify this, the voxel-based data for one of the subjects were down-sampled to
larger voxel sizes ($6\times 6\times 6$mm and $12\times 12\times 12$mm)
and the corresponding voxel-based networks were formed at these coarser resolutions.
FIG \ref{fig:scale} shows the degree distributions for voxel-based networks at different voxel
resolutions, as well as that of the region-based network. They all follow exponentially 
truncated power law distributions.
Table 1 shows the parameter estimates of $\beta$ and $k_c$ of 
$P(k)\propto k^{\beta} \exp(-k/ k_c)$ at different network scales, as well as the number
of nodes and the average node degree. As the network resolution increases from a coarse macroscopic 
representation of the region-based network to a finer mesoscopic representation of 
the voxel-based network, the exponent $\beta$ decreases
from positive to negative. The cut--off parameter $k_c$ on the other hand increases
as the number of nodes increases in finer representations of the network.
Our finding with the $12\times 12\times 12$mm scale network is consistent with
a previous study of region-based networks with 1,000 nodes demonstrating a clear 
characteristic of positive $\beta$ in the distribution (see Figure S1 of \cite{hagmann:core}).

\begin{table}
\caption{The estimated exponent $\beta$ and cut-off parameter $k_c$ 
for the exponentially truncated power law distributions for the networks 
with various scales, along with the number of nodes $N$ and the 
average node degree $\langle k \rangle$.}
\label{tab:table1}
\begin{ruledtabular}
\begin{tabular}{llcccc}
\multicolumn{2}{l}{Network} & $N$ & $\langle k \rangle$ & $\beta$ & $k_c$ \\
\hline
\multicolumn{2}{l}{Voxel-based network} \\
 & $4 \times 4 \times 5$mm voxel & 21585 & 29.4 & $-0.54$ & 63.3 \\
 & $6 \times 6 \times 6$mm voxel & 7727 & 20.1 & $-0.46$ & 37.4 \\
 & $12 \times 12 \times 12$mm voxel & 980 & 9.97 & 0.42 & 7.01 \\
\multicolumn{2}{l}{Region-based network} & 98 & 4.49 & 1.93 & 1.54 \\
\end{tabular}
\end{ruledtabular}
\end{table}

Our results above indicate that the scale at which the network is formed is important.
The higher the resolution, the smaller $\beta$ becomes, 
resembling more closely a power law distribution $P(k)\propto k^{-\gamma}$. 
This means that the network resembles scale-free networks more as its scale becomes finer.
This may be because connections and edges that are represented by a single edge 
or a single node in a coarse network can be represented with multiple edges 
and nodes in a finer network. A single low degree node of a coarse network 
may be represented by distinct low degree nodes in a finer network. 
Similarly, the edges at a hub in a coarse 
network may be resolved as collections of edges in a finer network, increasing the degree at the 
hub dramatically. A large number of low degree nodes and a very small number of
extremely high degree hubs are characteristics of scale-free networks 
\cite{barabasi:scalefree}. It is important to note that $\beta$ in an exponentially truncated 
power law distribution does not become smaller than $-1$ as in a power law distribution
of a true scale-free network $P(k)\propto k^{-\gamma}$, $\gamma>0$. 
However, it is possible to approximate a power law distribution by modeling the 
distribution with two curves: $P(k)\propto k^{-\gamma}$ for $k\leq k_c$ 
and $P(k)\propto k^{-\gamma} \exp(-k/ k_c)$ for $k>k_c$ \cite{grove,mossa}.
In such a parameterization, the tail above the cut-off
parameter $k_c$ is modeled as an exponentially truncated power law distribution
showing accelerated decay, but the distribution of the vast majority of nodes, nodes with 
degree $k \leq  k_c$, is modeled by a power law distribution.

%
%
It is interesting to note that exponentially truncated power law distributions have
been observed not only in the brain network but also other types of networks, including
ecological \cite{lusseau}, social \cite{newman:collab:i,amaral:pnas}, and technological
\cite{amaral:pnas,mossa} networks. Such networks may occur due to limitations in 
processing capability of nodes or restrictions in network growth. 
Mossa {\it et al.} \cite{mossa} generated networks
with information filtering, or a limitation on nodes' information processing capability,
and observed a truncation in the node degree distribution. Similarly, Amaral {\it et al.}
\cite{amaral:pnas} speculated that the cost of additional connections to a node
functioning near its capacity may limit the occurrence of mega-hubs commonly seen
in a scale-free network. A scale-free network can be formed as a result of a process known
as preferential attachments \cite{barabasi:scalefree}, in which a new node introduced
to a network preferentially attaches itself to a high-degree node, resulting a very small
number of mega hubs. However this preferential attachment process may be realistic
only when the network growth is unrestricted. Such a condition is not feasible for 
the brain network since there are physiological, anatomical, and functional constraints.
However, there may be some advantages for the brain network not being scale-free.
A scale-free network is an ideal structure for efficiently transmitting information to the
entire network, but this also means unnecessary or unwanted information can be
spread throughout the network very easily. Such an uncontrollable spread has been
observed in epidemiological networks in which a disease spreads to the entire network 
without any threshold for epidemics  \cite{pastor:vespignani,boguna}. 
On the other hand, an exponentially truncated power law network is resistant
against such massive epidemics, with a non-zero threshold for an epidemic
\cite{mossa}.  In the context of the brain network, a scale-free structure could cause 
extensive synchronization often seen in epilepsy patients, while an exponentially 
truncated network may prevent such over synchronization. This may also allow
localized processing in a certain module of the brain without involving the entire
brain network unless the demand of a cognitive process exceeds a certain threshold.

%
%
In summary, degree distributions of macroscopic and mesoscopic functional
brain networks were examined, through constructing region-based and voxel-based 
networks, respectively, on the same subjects. It was found that the degree distributions of
both networks follow exponentially truncated power law distributions, and not
power law distributions as previously reported on voxel-based networks.
The discrepancy in the literature is likely due to the fact that prior studies of voxel-based
networks plotted the degree distributions $P(k)$ whereas all region-based network
analyses plotted the cumulative distribution $1-F(k)$. Due to sensitivity to noise in the
extreme tail of the degree distribution, the cumulative distribution plots are preferred.
It was also found that the distributions of region-based and voxel-based networks
belong to the same continuum of
exponentially truncated networks, and finer scales of the network result
in networks more similar to scale-free networks. The results from this study suggest
the importance of modeling the brain network at resolutions as fine as data would
allow.

%
%
This work was supported by the Translational Scholar Award to S. H from the 
Translational Science Institute of Wake Forest University. Data collection for this
work was supported by the National Institute of Health (NS042658), as well as
the Roena Kulynych Memory and Cognition Research Center and the
General Clinical Research Center (RR07122) of Wake Forest University.
We would like to thank Drs. Ann Peiffer and Christina Hugenschmidt for collecting
the data.

%
%
\bibliography{./swn}

\end{document}